\documentclass{llncs}
\usepackage{makeidx}  
\usepackage{graphicx}
\usepackage{amsmath,amssymb} 
\newcommand{\ket}[1]{ \left |#1\right \rangle}
\makeindex

\begin{document}

\frontmatter          
\setcounter{page}{1}

\pagestyle{headings}  

\title{Measurement-Based Quantum Turing Machines and their Universality}
\author{Simon Perdrix, Philippe Jorrand}
\institute{Leibniz Laboratory\\ 46, avenue F\'elix Viallet 38000 Grenoble, France\\ \emph{simon.perdrix@imag.fr, philippe.jorrand@imag.fr}}
\maketitle
\begin{abstract}

Quantum measurement is universal for quantum computation (Nielsen \cite{N01}, Leung \cite{L01,L03}, Raussendorf \cite{R00,R03}). This universality allows alternative schemes to the traditional three-step organisation of quantum computation: initial state preparation, unitary transformation, measurement. In order to formalize these other forms of computation, while pointing out the role and the necessity of classical control in measu\-rement-based computation, and for establishing a new upper bound of the minimal resources needed to quantum universality, a formal model is introduced by means of Measurement-based Quantum Turing Machines.

\end{abstract}

\section{Introduction}
%%%%%%%

The driving force of research in quantum computation \cite{KSV,NC} is that of looking for the consequences of having information encoding, processing and communication make use of quantum physics, i.e. of the ultimate knowledge that we have, today, of the physical world, as described by quantum mechanics. Quantum mechanics, the mathematical formulation of quantum physics, relies on four postulates: (i) the state space of a quantum system is a Hilbert space; (ii) the evolution of the state of a closed quantum system is deterministic and characterized by a unitary operator; (iii) measurement, i.e. the evolution of a quantum system interacting with its (classical) environment is probabilistic and characterized by an hermitian operator named observable; and (iv) the state space of a quantum system composed of several quantum subsystems is the tensor product of the state spaces of its components. The question is then: how to take advantage of these postulates to the benefits of computation?

The most common approach to quantum computation exploits all four postulates in a rather straightforward manner. The elementary carrier of information is a qubit: the state of a $n$-qubit register lives in a $2^n$-dimensional Hilbert space, the tensor product of $n$ $2$-dimensional Hilbert spaces (postulates i and iv). Then, by reproducing in the quantum world the most traditional organization of classical computation, quantum computations are considered as comprising three steps in sequence: first, initial state preparation (postulate iii can be used for that, possibly with postulate ii); second, computation by deterministic unitary state transformation (postulate ii); and third, output of a result by probabilistic measurement (postulate iii).

The second step assumes that the $n$-qubit register is a closed quantum system, i.e. does not interact with its environment while the computation is going on. This creates very severe difficulties for the implementation of physical quantum computing devices. A physical qubit is indeed necessarily interacting with an external physical environment, either because the qubit is constrained to reside in some precise location (e.g. ion traps), or because it ÒfliesÓ across some free space while being operated upon (photons). In both cases, the state becomes entangled with (i.e. dependent upon) the states of particles belonging to the environment: the state is altered and, after a time depending on the chosen technology, it is no longer relevant for the ongoing computation. This unavoidable physical process, known as decoherence, can be kept under control by means of quantum error correcting codes. But this is highly resource consuming. Depending on the scheme chosen, such codes require 3 to 9 physical qubits per logical qubit, and the correction process takes time ($10^4$ to $10^6$  elementary unitary operations must be achievable within the decoherence time for ensuring successful error correction). 

Then, the idea is: instead of trying to climb over the high and steep physical obstacle of decoherence, just avoid it, disregard the second postulate, and rely upon the three other postulates only. In addition to being physically motivated, this has also been proved to be computationally relevant. Nielsen \cite{N01} has indeed shown that a generalization of quantum teleportation can be used for designing a universal quantum computation scheme based on measurement on at most 4 qubits. Leung \cite{L01,L03} has improved this result by showing that measurements on at most 2 qubits are universal. A seemingly very different approach has been proposed by Briegel and Raussendorf \cite{R00,R03}. In their "One-way quantum computer", a grid of qubits is initially prepared in a special fully entangled state, the "cluster state", where some of the qubits encode the input of the computation and others are designated as output qubits. Computation then operates stepwise, by successively measuring individual qubits: at each step, a yet unmeasured qubit and an observable are chosen and the corresponding measurement is applied. While the initial entanglement is consumed step by step by these measurements, a result is eventually pushed to the output qubits. However these two measurement-based quantum computation schemes, both proved to be universal, are still rather specific. There is a need for a more abstract model, of which both would be instances.

In this paper we introduce a new family of abstract quantum computation models, Measurement-based Quantum Turing Machines (MQTM). A hierarchy of models of that sort, ranked according to the amount of resources they use, are proved to be universal for quantum computation. While also pointing out the necessity and the precise role of classical control in measurement-based quantum computation, one of these models exhibits a new upper bound for the minimal resources needed for achieving quantum universality. Another class of such models, with slightly more restricted resources, is proved universal for classical computations, and proved not universal for quantum computations, thus characterizing, by comparing the amounts of resources, where the gap is between the classical and the quantum approaches to computing.

\section{Notations}

We briefly introduce notations which are used for describing quantum states, unitary transformations and quantum measurements, referring to the books by Nielsen and Chuang \cite{NC}, and by Kitaev, Shen and Vyalyi \cite{KSV} for more details. Here, a quantum system is a register of qubits. The state of a $n$-qubit register is a normalized vector in a $2^n$ dimensional Hilbert space $\mathcal{H}_n$. $\mathcal{B}=\{\ket{i}\}_{i\in \{0,1\}^n}$, where $\ket{i}$ is a vector in Dirac's notation, is the computational basis of $\mathcal{H}_n $. Thus a $n$-qubit state $\ket{\phi}$ in the computational basis is: $\ket{\phi}=\Sigma_{i\in\{0,1\}^n}\alpha_i\ket{i}$, $\alpha_i \in \mathbb{C}$. 

A unitary evolution $\mathcal{U}$, which acts on $n$ qubits, is represented by a $2^n\times 2^n$ unitary matrix $U$. $\mathcal{U}$ transforms $\ket{\phi}$ into $U\ket{\phi}$. Clearly, a unitary evolution is deterministic.

A measurement is represented by an observable $O$ which is a hermitian matrix. Considering the spectral decomposition of $O$, $O=\Sigma_mmP_m$, a $O$-measurement transforms, with probability $p_m$, a state $\ket{\phi}$ into the state $\frac{P_m\ket{\phi}}{\sqrt{p_m}}$, where $p_m$ is the scalar product of $\ket{\phi}$ and $P_m\ket{\phi}$. The classical outcome of the measurement is $m$. Thus a measurement is a probabilistic transformation of the state of the register, which returns a classical outcome $m$. The classical outcome gives information on the state of the system after the measurement.

Pauli matrices, $I,X,Y,Z$ are unitary matrices which can be viewed as unitary operators (in this case the notation $\sigma_x,\sigma_y,\sigma_z$ is prefered to $X,Y,Z$). They can also be viewed as observables since they are hermitian. Pauli matrices form a group (up to a global phase), e.g. $X^2=I$ and $X.Y=i.Z$. A $Z$-measurement is also called a measurement in the computational basis.

\begin{center}
$I=\left(\begin{array}{cc}
  1 & 0\\
  0&1\\
\end{array} \right)$, 
$X=\sigma_x=\left(\begin{array}{cc}
  0 & 1\\
  1&0\\
\end{array} \right)$, 
$Y=\sigma_y=\left(\begin{array}{cc}
  0 & -i\\
  i&0\\
\end{array} \right)$, 
$Z=\sigma_z=\left(\begin{array}{cc}
  1 & 0\\
  0&-1\\
\end{array} \right)$
\end{center}

%%%%%%
\section{Universalities}

%%%%%%
\subsection{Definitions}

There exist different abstract models for classical computation (e.g. Turing Machines, Automata) and for quantum computation (e.g. Quantum Circuits \cite{Y93,KSV}, Quantum Turing Machines \cite{D85,BV97}). We introduce in this section a way of comparing such models with respect to their power. The most powerful models are considered \emph{universal}.

A model is a set of machines. The power of models depends on the \emph{context} of their utilization. Only machines which transform an input $in$ into an output $out$ are considered. Contexts are pairs $In \times Out$ of a set $In$ of inputs and a set $Out$ of outputs. Each machine $m$ is represented by its action, i.e. a relation $m \in \mathcal{P}(In \times Out)$ ($\mathcal{P}(E)$ denotes the set of the subsets of $E$). For all $in \in dom(m)$, where $dom(m)\subset In$ is the domain of definition of $m$, $m(in)\subset Out$ represents the elements of $Out$ which are in relation with $in$ by $m$. In the deterministic case, $m$ is a function, so $m(in)\in Out$.

A machine is an element of $\mathcal{P}(In \times Out)$ (i.e. a subset of $In \times Out$), but not all the elements of $\mathcal{P}(In \times Out)$ are machines, because a machine must be \emph{physically realizable} or, in other words, the relation must be computable.
A model $Mod$ is a set of machines, so $Mod \in \mathcal{P}(\mathcal{P}(In \times Out))$. $\mathcal{R} \subset \mathcal{P}(\mathcal{P}(In \times Out))$ is the set of realizable models, i.e. models comprising physically realizable machines only.

In order to compare the power of two models, a notion of \emph{simulation} is introduced. A machine $m_1$ $S_1$-simulates a machine $m_2$ ($m_2 \prec_{S_1} m_1$) under the context $In \times Out$ iff for all $in \in dom(m_2)$, $m_1(in)=m_2(in)$. Simulation $\prec_{S_1}$ is a quasi-order relation on $\mathcal{P}(In \times Out)$.

In order to obtain a more restrictive simulation $\prec_{S_2}$ (i.e. $\prec_{S_2} \subset \prec_{S_1}$), some additional conditions of complexity may be introduced: a machine $m_1$ $S_2$-simulates a machine $m_2$ under the context $In \times Out$ iff $m_2 \prec_{S_1} m_1$ and $\forall in \in dom(m_2)$, $\tau_{m_1}(in)=O(\tau_{m_2}(in))$, where $\tau_m(in)$ is the execution time of $m$ on $in$. Moreover in order to obtain a less restrictive simulation $\prec_{S_3}$ (i.e. $\prec_{S_1} \subset \prec_{S_3}$), only some conditions of correction may be required: a machine $m_1$ $S_3$-simulates a machine $m_2$ under the context $In \times Out$ iff for all $in \in dom(m_2)$, $m_1(in)\subset m_2(in)$. 

In general, any quasi-order relation $\prec_S$ on $\mathcal{P}(In \times Out)$ is a simulation. This relation $\prec_S$ is extended to models: a model $Mod_1$ $S$-simulates a model $Mod_2$ ($Mod_2 \prec_{S} Mod_1$) under the context $In \times Out$ iff $\forall m_2 \in Mod_2, \exists m_1 \in Mod_1$ such that $m_2 \prec_S m_1$.

For a given context, a model $Mod_1$ is $S$-universal, iff $\forall Mod \in \mathcal{R}, Mod\prec_S Mod_1$.

For a given context, a given model $Mod$ and a given simulation $\prec_S$, $m_1\in Mod$ is a $Mod$-universal machine iff $\forall m \in Mod, \exists in_{m}\in In$ such that $m \prec_S m_1[in_m]$, where $m_1[in_m]$ is $m_1$ with $in_m$ always inserted in the input.

\subsection{Classical Universality}

The context of classical universality is a context of language recognition, i.e. for a finite vocabulary $V$, $In=V^*$ and $Out=\{true, false\}$. Note that the context $In=Out=V^*$ may also be chosen. A machine $m_1$ classically simulates $m_2$ ($m_2\prec_{Class} m_1$), iff for all $in \in dom(m_2), m_1(in)=m_2(in)$.

If $Mod_{TM}$ is the model of classical Turing Machines, the Church-Turing thesis is nothing but: $\forall Mod \in \mathcal{R}, Mod \prec_{Class} Mod_{TM}$. Thus a model $Mod$ is classically universal iff $Mod_{TM}\prec_{Class} Mod$.

Moreover, there exists a machine $m_{univ}\in Mod_{TM}$ which is universal, i.e. $\forall m \in Mod_{TM}, \exists in_{m} \in V^*$ such that $\forall in \in V^*, m \prec_{Class} m_{univ}[in_m]$

\subsection{Quantum Universality}

The context of quantum universality is a context of quantum states transformation, i.e. $In=\mathcal{H}$ and $Out=\mathcal{H}$, where $\mathcal{H}$ is the Hilbert space of quantum states. Note that the context of density matrices transformation may also be chosen. A machine $m_1$ quantum simulates $m_2$ ($m_2\prec_{Quant} m_1$), iff for all $in \in dom(m_2), m_1(in)=m_2(in)$.
 %%%%%%

 %%%%%%%
 \section{Measurement-based Quantum Turing Machines}

\subsection{Turing machine: from classical to quantum}

Because of the essential role of Turing machines ($TM$) in classical computer science, it is natural to consider a quantum analogue of them. A quantum Turing machine ($QTM$) \cite{D85,BV97} is an abstract model of quantum computers, which expands the classical model of a Turing machine by allowing a \emph{quantum} transition function: in a $QTM$, superpositions and interferences of configurations are allowed, but inputs and outputs of the machine are still classical. Thus the model of $QTM$ explores the computational power of quantum mechanics for solving classical problems, without considering \emph{quantum} problems, i.e. quantum input/output. 

While quantum circuits and quantum random access machines are mainly used to describe specific algorithms, the development of complexity classes, like $QMA$ \cite{W00}, which deal with quantum states, points out the necessity of theoretical models of quantum computation acting on quantum data.

In measurement-based quantum computations, classical conditional structures are required for controlling the computation. The classical control may be described as follows: \emph{"if the classical outcome of measurement number $i$ is $\lambda$, then measurement number $i+1$ is according to observable $O_a$, else measurement number $i+1$ is according to observable $O_b$"}. This classical control, which is also used in the scheme of quantum teleportation \cite{B93}, is essential for measurement-based quantum computation, but has never been formalized.

Therefore, a measurement-based quantum Turing machine ($MQTM$) is a $TM$ with a quantum tape for acting on quantum data, and a classical transition function for a formalized classical control. The heads of a $MQTM$, which are the natural connection between the classical control and the quantum tape of the machine, work according to the measurement postulate of quantum mechanics.

%%%%%%

 \subsection{Definitions}
 
  A $MQTM$ is composed of: $p\in {\mathbb{N}^*}$ tapes (finite or infinite) of \emph{quantum} cells (a quantum cell can be seen as a $d$-level particle, for convenience we consider $d=2$ so each quantum cell is nothing but a qubit) and $k\in {\mathbb{N}^*}$ measurement heads.

 \begin{definition}A $MQTM$ $M$ is defined by $( Q ,
\Sigma,\mathcal{O},\delta)$ where: $ Q $ is a finite set of classical states with an identified initial state $q_0$ and final state $q_f \neq q_0$, $
\Sigma$ is a finite alphabet of classical outcomes such that $\vert \Sigma \vert = 2^k$, $\mathcal{O}$ is a set of $k$-qubit observables (i.e. each measurement is on at most $k$ quantum cells, those which are under the measurement heads) such that all possible classical outcomes of each observable of $\mathcal{O}$ are in $
\Sigma$, and $\delta$ is a classical transition function $\delta :  Q \times 
\Sigma \to  Q  \times \mathcal{O} \times \mathcal{D}$, where $\mathcal{D}$ is a set of allowed movements of the heads. \end{definition} 

%%%%%%
 
 A configuration of the $MQTM$ is a complete description of the contents of the tapes (i.e. the quantum state $\ket{\psi} \in \mathcal{H}$ of the \emph{finite} quantum system composed of input cells and already visited cells), the location $l\in {\mathbb{Z}^k}$ of the tape heads, the last observed classical outcome $\lambda\in 
\Sigma$ (a given $\lambda_0\in 
\Sigma$ plays the role of the last outcome for the initial configuration) and the classical state $q\in  Q $. 
 
 A computation on a $MQTM$ $m$ operates as follows:
 \begin{itemize}
 \item Initially:
 \begin{itemize}
 \item The input of the computation $\ket{\varphi}$ (in principle, $\ket{\varphi}$ is unknown) is placed on a specified tape, all others qubits are in an unknown state, but not entangled with each other nor with the input qubits.
 \item A specified measurement head points on the first quantum cell of the input.
 \item The classical state of $m$ is $q_0$ and $\lambda_0$ plays the role of the last measurement outcome. 
 \end{itemize}
 
 \item Then, transitions are successively applied, transforming the quantum state $\ket{\psi}$ of the tapes.
 \item Computation halts when the classical state of the machine is $q_f\in  Q $. At that time, a specified head points on the quantum output of the computation. 
 \end{itemize}

One may notice that transitions permit a formal description of communications between the classical and quantum worlds. Like in a classical $TM$, $\delta$ is nothing but the \emph{program} of the machine. Since $Q \times \Sigma$ is finite, the image by $\delta$ of $Q \times \Sigma$ is also finite, therefore only a finite part of $\mathcal{O}$ may be used. This limitation to a finite subset of $\mathcal{O}$ is a consequence of the classical control.

 $M_{MQTM}$ is the model comprising all Measurement-based Quantum Turing Machines operating according to this principle.

 \subsection{Universal models of $MQTM$}
 
 In the following, a model $M$ is characterized by its resources: number of tapes, number of measurement heads, allowed movements for the heads and set of observables.

In the model of quantum computation by measurement improved by Leung \cite{L01}, the set of two-qubit observables $\{ X\otimes X, Z\otimes Z, X\otimes Z,Z\otimes X, X\otimes I,Z\otimes I, \frac{1}{\sqrt{2}}(X\otimes X+Y\otimes X)  \}$ is proved to be quantum universal. This leads to the following lemma, which is proved in \cite{PJ04}.

\begin{lemma}
The model $M_A\subset M_{MQTM}$ of $MQTM$ composed of one infinite tape, two measurement heads, $\mathcal{D}_A=\mathbb{Z}^2$ and $\mathcal{O}_A=\{ X\otimes X, Z\otimes Z, X\otimes Z,Z\otimes X, X\otimes I,Z\otimes I, \frac{1}{\sqrt{2}}(X\otimes X+Y\otimes X),\frac{1}{\sqrt{2}}(X\otimes X+X\otimes Y)    \}$, is quantum universal.
\end{lemma}
\begin{center}
\includegraphics[width=0.4\textwidth]{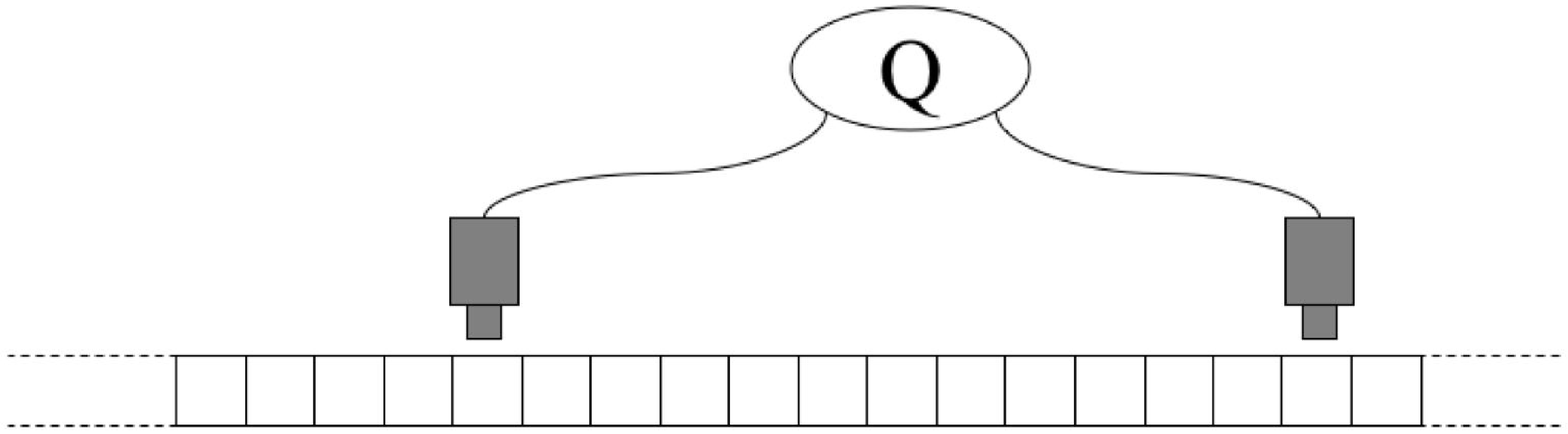}

\emph{\small Figure 1. Machines in model $M_A$}
\end{center}
 
 \emph{Lemma 1} is nothing but the translation of the universality results proved by Leung \cite{L01} into the formalism of $MQTM$. The main point of the proof of \emph{Lemma 1} is to ensure that, even if some executions never terminate, the set of classical states $Q_{m_A}$ is \emph{finite}, for all $m_A\in M_A$.
 
We now introduce models $M$ of $MQTM$ with less resources than $M_A$ (i.e. such that the simulation $M\prec_{Quant} M_A$ is trivial) in order to exhibit a hierarchy of models of $MQTM$, eventually drawing the frontier between universal and non-universal models.

 \begin{lemma}
 For any set $\mathcal{O}_B$ of $1$-qubit observables and for any $\mathcal{D}_B \subset \mathbb{Z}$ , the model $M_B\subset M_{MQTM}$ of $MQTM$ composed of one infinite tape, one measurement head, $\mathcal{D}_B$ and $ \mathcal{O}_B$, is \emph{not} quantum universal.  
\end{lemma}
 
 \begin{proof}
By counter-example. Entanglement can not be created using only one-qubit measurements, i.e. for a given register of two qubits in the state $\ket{00}$ (which is a separable state) and for any sequence $s$ of one-qubit measurements, the state of the register after the application of $s$ is separable. Thus the unitary transformation $U=(H\otimes I).CNot$, which transforms the separable state $\ket{00}$ into the entangled state $(\ket{00}+\ket{11})/\sqrt{2}$ can not be simulated by a machine of $M_B$, whereas there exists $m_A\in M_A$ which simulates $U$. So $M_B$ is not quantum universal.
 $\hfill \Box$
 \end{proof}
 
 \begin{center}
\includegraphics[width=0.4\textwidth]{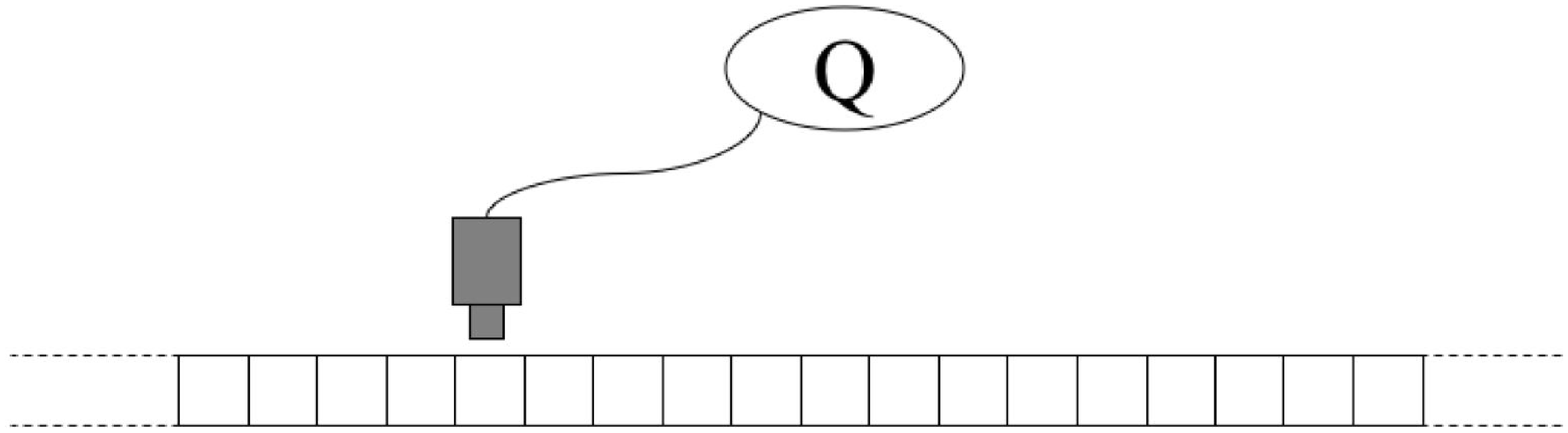}

\emph{\small Figure 2. Machines in models $M_B$ and $M_C$}
\end{center}

One may wonder why $M_B$ is not quantum universal whereas Briegel and Raussendorf have proved, with their One-way quantum computer, that one-qubit measurements are universal \cite{R00,R03}. The proof by Briegel and Raussendorf is given with a strong assumption which is that there exists a grid of auxiliary qubits which have been initially prepared, by some unspecified external device, in a globally entangled state (the cluster state), whereas \emph{creation} of entanglement is a crucial point in the proof of \emph{Lemma 2}. Moreover, another strong assumption of one-way quantum computation is that the input state $\ket{\varphi}$ has to be classically known (i.e. a mathematical description of $\ket{\varphi}$ is needed), whereas the manipulation of unknown states (i.e. manipulation of qubits in an unknown state) is usual in quantum computation (e.g. teleportation \cite{B93}). Since none of these assumptions are verified by $M_B$, \emph{Lemma 2} does not contradict the results of Briegel and Raussendorf.

\begin{lemma}
The model $M_C\subset M_{MQTM}$ of $MQTM$ composed of one infinite tape, one measurement head, $\mathcal{D}_C=\{-1,0,+1\}$ and $\mathcal{O}_C=\{ X, Z\}$, is classically universal, i.e. $M_{TM} \prec_{Class} M_C$.
\end{lemma}

 \begin{proof} We prove that $M_C$ classically simulates the model $M_{TM}$ of classical Turing Machines composed of an infinite tape of bits and a read-write head, i.e. for any classical Turing Machine $m\in M_{TM}$, there exists a machine $m_C\in M_C$, such that $m\prec_{Class} m_C$. For a given $m\in M_{TM}$, a machine $m_C \in M_C$ is considered, such that: the tape of qubits plays the role of the tape of bits; the classical value $0$ (resp $1$) is represented by the quantum state $\ket{0}$ (resp $\ket{1}$). In order to simulate classical reading, a $Z$-measurement is performed: if the state of the qubit is $\ket{0}$ the classical outcome of the measurement is $0$ with probability one, same for $\ket{1}$ with the classical outcome $1$. Thus the measurement head of $m_C$ plays the role of the reading head of $m$. In order to simulate writing, for instance of the value $0$ on a bit, an $X$-measurement followed by a $Z$-measurement are performed. After the $X$-measurement, the state of the qubit is $(\ket{0}+\ket{1})/\sqrt{2}$ or $(\ket{0}-\ket{1})/\sqrt{2}$, and, after the $Z$-measurement, is $\ket{0}$ with probability $1/2$, and $\ket{1}$ with probability $1/2$. If the state is $\ket{1}$, the process ($X$-measurement followed by $Z$-measurement) is repeated, until it becomes $\ket{0}$.
 $\hfill \Box$
 \end{proof}
 
 $M_C$ is classically universal (\emph{Lemma 3}), but $M_C$ is not quantum universal (\emph{Lemma 2}), so this model points out a gap between classical computation and quantum computation. From a decidability point of view, quantum and classical computations are equivalent \cite{D85}, which seemed to imply that the only difference concerns complexity issues. However, in our definitions for quantum and classical universalities, there are no restrictions on complexity, therefore one may wonder why we find this gap between quantum and classical universalities. The key is that contexts of utilization differ: for classical universality, machines act on classical inputs, for quantum universality, machines act on quantum inputs.

\begin{lemma}
The model $M_D\subset M_{MQTM}$ of $MQTM$ composed of two infinite tapes, one measurement head per tape, $\mathcal{D}_D=\mathbb{Z}^2$ and $\mathcal{O}_D=\{ X\otimes X, Z\otimes Z, X\otimes Z,Z\otimes X, X\otimes I,Z\otimes I, I\otimes X,I\otimes Z, \frac{1}{\sqrt{2}}(X\otimes X+X\otimes Y),\frac{1}{\sqrt{2}}(X\otimes X+Y\otimes X)   \}$, is quantum universal, i.e. $M_A \prec_{Quant} M_D$.
\end{lemma}
\begin{center}
\includegraphics[width=0.4\textwidth]{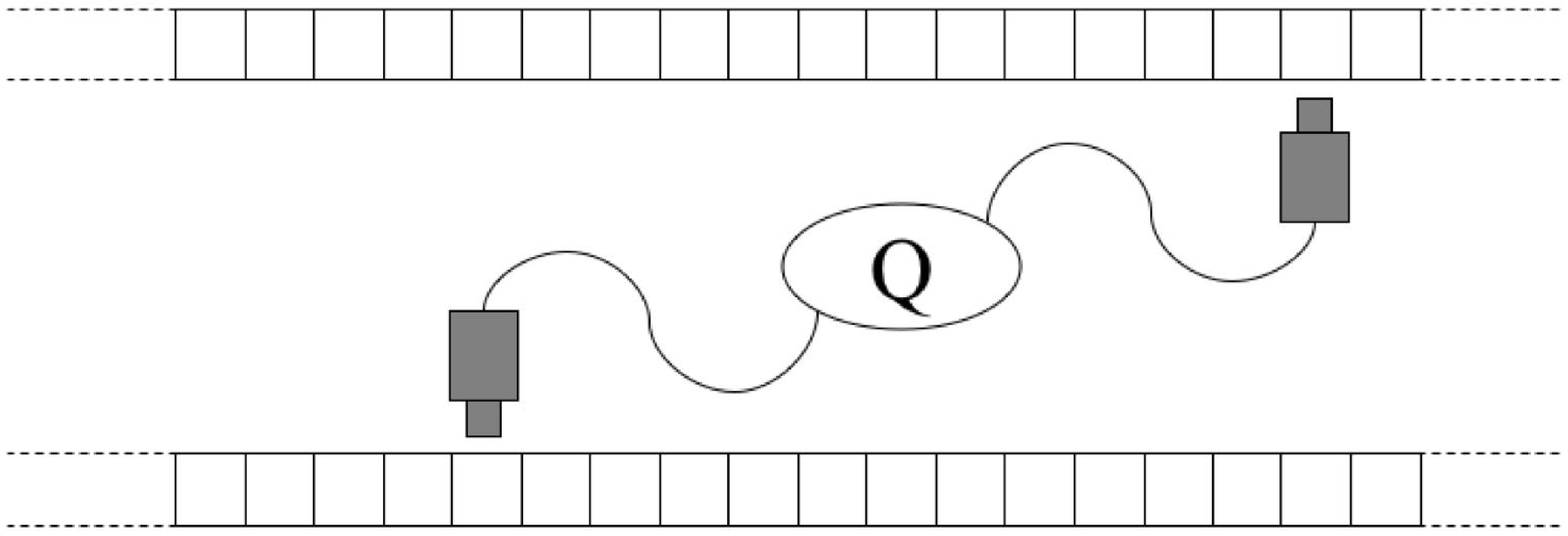}

\emph{\small Figure 3. Machines in model $M_D$}

\end{center}
\begin{proof}
 We prove that $M_D$ quantum simulates $M_A$, i.e. for any machine $m_A\in M_A$, there exists a machine $m_D\in M_D$, such that $m_A\prec_{Quant} m_D$.
Qubits of $m_A$ are indexed from $-\infty$ to $+\infty$.
The machine $m_D$ has two tapes: its upper tape and its lower tape (see fig.3). A subset of the qubits of the upper tape of $m_D$ are indexed from $-\infty$ to $+\infty$, using odd numbers only, while a subset of the qubits of the lower tape are numbered with even numbers only, such that there remain an infinite number of non-indexed qubits on each tape of $m_D$. These non-indexed qubits will be available as \emph{auxiliary qubits}.

An execution on $m_A$ is entirely described by a sequence of measurements. We show that each $O$-measurement in this sequence may be simulated by $m_D$. A two-qubit $O$-measurement of $m_A$ acts on qubits of $m_A$ indexed by $i$ and $j$.
\begin{itemize}
\item If $i$ and $j$ have a different parity, a $O$-measurement on qubits $i$ and $j$ is allowed on $m_D$, because $i$ and $j$ are not on the same tape and $O \in \mathcal{O}_D$.
\item Otherwise, assume $i$ and $j$ are both even (so $i$ and $j$ are on the lower tape of $m_D$, see fig. 4). If the state of $j$ is teleported \cite{B93} from the lower tape to an auxiliary qubit $a$ of the upper tape, then $O$ may be applied on $i$ and $a$. A second teleportation from $a$ to $j$ will then terminate the simulation of a $O$-measurement on $i$ and $j$. Thus, the quantum universality of $M_D$ is reduced to the ability to teleport the state of a qubit from one tape of $m_D$ to the other.
\begin{center}
\includegraphics[width=0.4\textwidth]{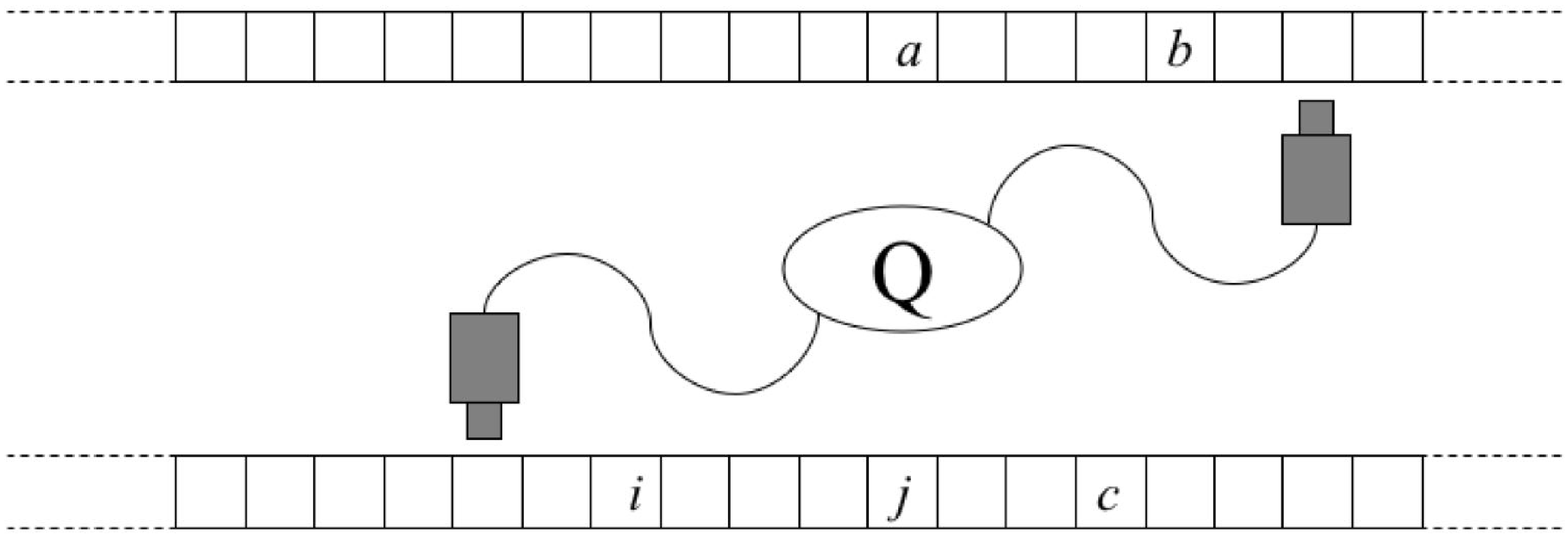}

\emph{\small Figure 4. Simulation of $m_A$ by $m_D$}

\end{center}
Considering a qubit $j$ of the lower tape, two auxiliary qubits $a$ and $b$ of the upper tape, and an auxiliary qubit $c$ of the lower tape, teleporting $j$ to $a$ consists in assigning a Bell state to $a$ and $b$, then performing a Bell measurement on $j$ and $b$. A Bell measurement may be decomposed into a $Z\otimes Z$-measurement followed by a $X\otimes X$-measurement. Applying a Bell measurement on $a$ and $b$ assigns a Bell state to these two qubits, but $a$ and $b$ are both on the upper tape, so a measurement on these two qubits cannot be performed, that is why an auxiliary qubit $c$ is needed. Using $c$, the sequence of measurements $\{I \otimes Z^{(a)},I \otimes Z^{(b)},Z^{(c)}\otimes I, X^{(c)} \otimes X^{(a)}, X^{(c)} \otimes X^{(b)},Z^{(c)} \otimes I\}$, assigns a Bell state to qubits $a$ and $b$. 

The state of the 3-qubit register $a,b,c$ after the first three measurements in this sequence is $\ket{\psi}=(\sigma_x^{\frac{1-i}{2}}\otimes \sigma_x^{\frac{1-j}{2}}\otimes \sigma_x^{\frac{1-k}{2}})\ket{000}$, where $i,j,k\in\{-1,1\}$ are the respective classical outcomes of these measurements. Then the evolution of $\ket{\psi}$ with the remaining measurements in the sequence is:

$\ket{\psi_1}=(\sigma_z^{\frac{1-l}{2}}\sigma_x^{\frac{1-i}{2}}\otimes \sigma_x^{\frac{1-j}{2}}\otimes  \sigma_x^{\frac{1-k}{2}})[\frac{1}{\sqrt{2}}(\ket{000}+\ket{101})]$

$\ket{\psi_2}=(\sigma_z^{\frac{1-l}{2}}\sigma_x^{\frac{1-i}{2}}\otimes \sigma_z^{\frac{1-m}{2}}\sigma_x^{\frac{1-j}{2}}\otimes \sigma_x^{\frac{1-k}{2}})[\frac{1}{2}(\ket{000}+\ket{011}+\ket{101}+\ket{110})]$

$\ket{\psi_3}=( \sigma_x^{\frac{1-k}{2}}\sigma_z^{\frac{1-l}{2}}\sigma_x^{\frac{1-i}{2}}\otimes  \sigma_x^{\frac{1-n}{2}}\sigma_z^{\frac{1-m}{2}}\sigma_x^{\frac{1-j}{2}}\otimes  \sigma_x^{\frac{1-n}{2}}\sigma_x^{\frac{1-k}{2}})[\frac{1}{\sqrt{2}}(\ket{00}+\ket{11})\otimes\ket{0}]$

where $l,m,n \in \{-1,1\}$ are the respective classical outcomes of the last three measurements. At the end, the state of $a$ and $b$ is a Bell state.

Because of the probabilistic aspect of quantum measurement, teleportation succeeds with probability $1/4$. If it does not succeed, a process of correction which consists in teleporting the state of $a$ to another auxiliary qubit, is repeated until a satisfactory state is produced.  $\hfill \Box$
\end{itemize}
\end{proof}

The previous proof is based on a reduction of \emph{Lemma 4} to the ability to teleport a state from a tape to the other. This teleportation is itself reduced to the ability to assign a Bell state to two qubits of the same tape.

Since the teleportation of a state from a tape to the other is possible, one can wonder if both tapes need to be infinite or if all qubits can be "stored" on the same tape and teleported on the other tape when needed. In this case how many cells on the finite tape are required for teleporting a state from a tape to the other? The answer is in the following lemma:

\begin{lemma}
The model $M_E\subset M_{MQTM}$ of $MQTM$ composed of a two-qubit tape, an infinite tape, one measurement head per tape, $\mathcal{D}_E=\{-1,0,1\}\times \mathbb{Z}$ and $\mathcal{O}_E=\{ X\otimes X, Z\otimes Z, X\otimes Z,Z\otimes X, X\otimes I,Z\otimes I, I\otimes X,I\otimes Z, \frac{1}{\sqrt{2}}(X\otimes X+X\otimes Y),\frac{1}{\sqrt{2}}(X\otimes X+Y\otimes X)   \}$, is quantum universal, i.e. $M_A \prec_{Quant} M_E$.
\end{lemma}
\begin{center}
\includegraphics[width=0.4\textwidth]{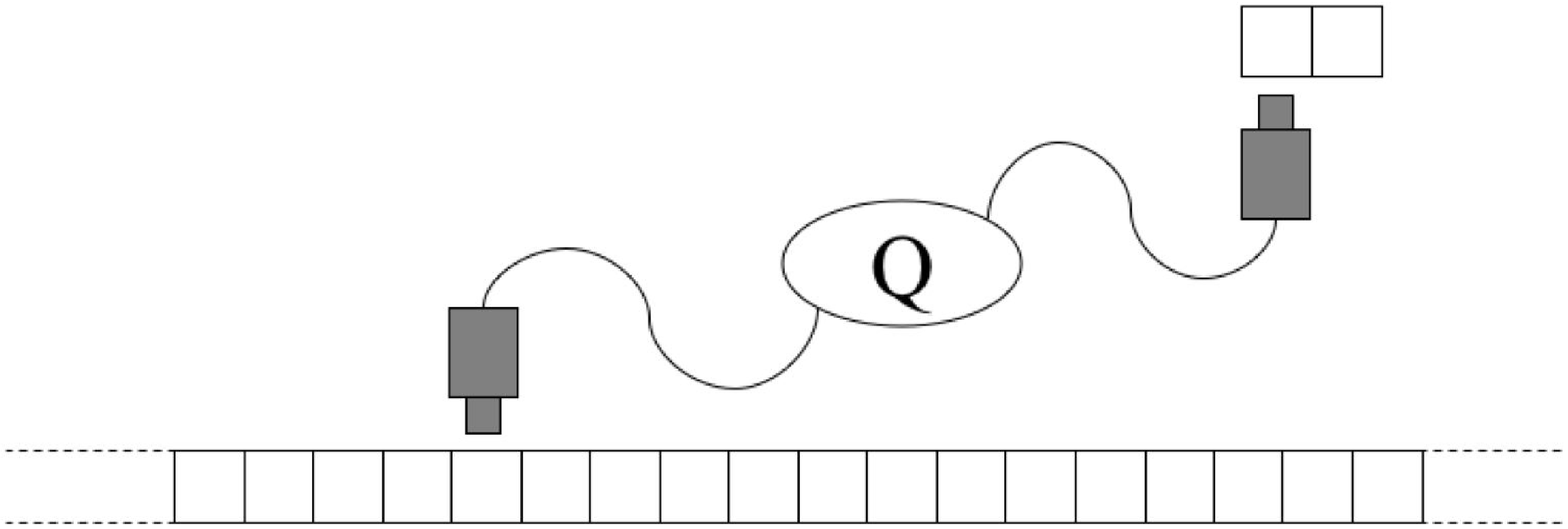}

\emph{\small Figure 5. Machines in model $M_E$}

\end{center}

\begin{proof} 
We prove that $M_E$ quantum simulates $M_A$, i.e. for any machine $m_A\in M_A$, there exists a machine $m_E\in M_E$, such that $m_A\prec_{Quant} m_E$.
Qubits of $m_A$ are indexed from $-\infty$ to $+\infty$. A subset of the qubits of the infinite tape of $m_E$ are indexed from $-\infty$ to $+\infty$ such that there remain an infinite number of non-indexed qubits on the infinite tape of $m_E$, which will be available as \emph{auxiliary qubits}.

An execution on $m_A$ is entirely described by a sequence of measurements. We show that each $O$-measurement in this sequence may be simulated by $m_E$. A two-qubit $O$-measurement acts on qubits of $m_A$ indexed by $i$ and $j$. Qubits of the finite tape of $m_E$ are indexed by $a$ and $b$. To simulate a $O$-measurement on $i$ and $j$, the state of $i$ is teleported to $a$, then $O$ is performed on $a$ and $j$, then the state of $a$ is teleported to $i$. Thus the quantum universality of $M_E$ is reduced to the ability to teleport a state from the infinite tape to the finite tape and vice-versa. In the proof of \emph{Lemma 4}, we have seen that the teleportation of a state from one tape to the other needs two qubits on each tape, including the teleported qubit. Thus $M_E$ quantum simulates $M_A$.$\hfill \Box$
\end{proof}

Since $M_E$ is universal (\emph{Lemma 5}), one may wonder if a more restrictive model with a one-cell instead of two-cell finite tape is universal. Using the teleportation scheme, a state of the infinite tape cannot be teleported to the one-qubit tape. This implies that the proof of the following theorem be based not on teleportation, but on a less resource consuming process known as \emph{state transfer}, which we define and use in \cite{P04,JP04} for studying measurement-based quantum computations. Note that a process equivalent to state transfer has also been used more recently \cite{AL04,CLN04} for the same purpose.

\begin{theorem}
The model $M_F\subset M_{MQTM}$ of $MQTM$ composed of a one-qubit tape, an infinite tape, one measurement head per tape, $\mathcal{D}_F=\{0\}\times \mathbb{Z}$ and $\mathcal{O}_F=\{ X\otimes X, Z\otimes Z, X\otimes Z,Z\otimes X, X\otimes I,Z\otimes I, I\otimes X,I\otimes Z, \frac{1}{\sqrt{2}}(X\otimes X+X\otimes Y)\}$, is quantum universal.
\end{theorem}

\begin{proof} Like in \emph{Lemma 5}, the proof is based on the ability to transfer the state of a qubit $j$ from the infinite tape to the qubit $a$ of the other tape and vice-versa. State transfer from $j$ to $a$ consists in $Z$-measuring $a$, then $X\otimes X$-measuring $a$ and $j$, then $Z$-measuring $j$ (see \cite{P04} for details). For a given state $\ket{\psi}=(\alpha\ket{0}+\beta \ket{1})\otimes (\gamma \ket{0} +\delta \ket{1})$ of the 2-qubit register $j,a$, the evolution of $\ket{\psi}$ along the sequence of measurements is:

$\ket{\psi_1}=(I\otimes \sigma_x^{\frac{1-i}{2}})[(\alpha \ket{0}+\beta \ket{1})\otimes \ket{0}]$

$\ket{\psi_2}=(I\otimes \sigma_z^{\frac{1-j}{2}}\sigma_x^{\frac{1-i}{2}})[\alpha \ket{00}+\beta \ket{01}+\beta \ket{10}+\alpha \ket{11}]$

$\ket{\psi_3}=(\sigma_x^{\frac{1-k}{2}}\otimes \sigma_x^{\frac{1-k}{2}}\sigma_z^{\frac{1-j}{2}}\sigma_x^{\frac{1-i}{2}})[\ket{0}\otimes (\alpha \ket{0}+\beta \ket{1})]$

where $i,j,k\in\{-1,1\}$ are the respective classical outcomes of the three successive measurements. In order to transfer a state from $a$ to a qubit of the infinite tape the same scheme is applied. The state transfer succeeds up to a Pauli operator. If this Pauli operator is not $I$ then a correcting process is performed, consisting in transferring the state again. $\hfill{\Box}$
\end{proof}

Comparing models $M_F$ and $M_C$, the minimal resources for quantum universality seem to be reached. An ultimate improvement is given in \emph{Theorem 2} by proving that if the movements of the heads are restricted to the natural 3 possible movements: \emph{one step left, no head movement, one step right}, then the model is still universal.

\begin{theorem}
The model $M_G\subset M_{MQTM}$ of $MQTM$ composed of a one-qubit tape, an infinite tape, one measurement head per tape, $\mathcal{D}_G=\{0\}\times \{-1,0,1\}$ and $\mathcal{O}_G=\{ X \otimes X, Z \otimes Z, X \otimes Z, Z \otimes X, X \otimes I, Z \otimes I, I \otimes X, I \otimes Z, \frac{1}{\sqrt{2}}(X\otimes X+X\otimes Y)\}$, is quantum universal.
\end{theorem}
 \begin{center}
\includegraphics[width=0.4\textwidth]{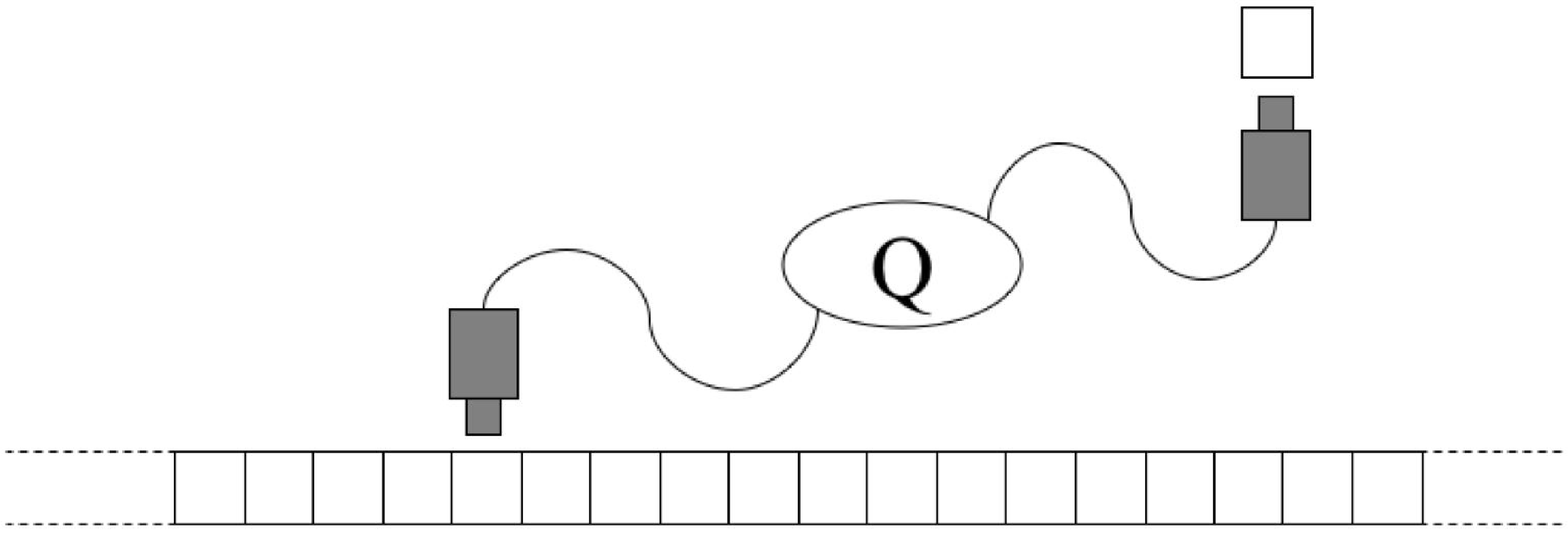}

\emph{\small Figure 6. Machines in the model $M_G$, with the minimal resources for quantum universality}

\end{center}
 \begin{proof}
 
 In order to prove that $M_G$ is quantum universal, we prove $M_{F}\prec_{Quant}M_G$ i.e. for any machine $m_F\in M_{F}$, there exists a machine $m_G\in M_G$, such that $m_F\prec_{Quant} m_G$. 

For a given $m_F \in M_F$, we consider $m_G\in M_G$ such that $\Sigma_{m_G}=\Sigma_{m_F}$, and for each $(q,\lambda)\in Q_{m_F}\times \Sigma_{m_F}$, if $\delta_{m_F}(q,\lambda)=(q',O,d)$ and $d=(0,\tau .k)$ with $\tau \in\{-1,1\}$ and $k\in \mathbb{N}$, then the states $q, q'$ and $q^{(0)},\ldots,q^{(k)}$ are in $ Q_{m_G}$, with the following transition function:

\begin{center}
\begin{tabular}{rrcl}
& $\delta_{m_G}(q,\lambda)$&$=$&$(q^{(0)},I\otimes I,(0,0))$\\
$\forall j\in\{1..k\},$&$\delta_{m_G}(q^{(j-1)},\_\ )$&$=$&$(q^{(j)},I\otimes I,(0,\tau))$\\
&$\delta_{m_G}(q^{(k)},\_\ )$&$=$&$(q',O,(0,0))$
\end{tabular}
\end{center}
A $I\otimes I$-measurement means that no measurement is done. One can notice that the set $ Q_{m_G}$ is finite. Thus $m_G$ quantum simulates $m_F$, so $M_G$ is quantum universal.$\hfill \Box$
\end{proof}

%%%%%%

\section{Conclusion}

This paper introduces a unified formalization of the concepts of classical and quantum universalities. This led to the introduction of a new abstract model for quantum computations, the model of Measurement-based Quantum Turing Machines ($M_{MQTM}$).

This model allows a rigorous formalization of the necessary interactions between the quantum world and the classical world during a measurement-based quantum computation. $M_{MQTM}$ has been studied within a general framework of abstract computation models, with a notion of simulation among models allowing to compare them in terms of their universalities.

Two main results have been obtained in this framework. A hierarchy of models contained in $M_{MQTM}$, ranked according to the resources they use, have been proved universal for quantum computations. One of them ($M_G$) exhibits a new upper bound for the minimal resources required for quantum universality.

Another subset of $M_{MQTM}$ ($M_C$), with more restricted resources, has been proved universal for classical computations, and proved not universal for quantum computations, thus pointing out a gap between the classical and quantum approaches to computing.

\end{document}